\pdfoutput=1

\documentclass[11pt]{article}

\usepackage[preprint]{acl}

\usepackage{times}
\usepackage{latexsym}

\usepackage[T1]{fontenc}

\usepackage[utf8]{inputenc}

\usepackage{microtype}

\usepackage{inconsolata}

\usepackage{graphicx}

\usepackage{graphicx}
\usepackage{amsmath}
\usepackage{amsfonts}
\usepackage{amsthm}
\usepackage{bbm}
\usepackage{hyperref}
\usepackage{tikz}
\usepackage{bm}
\usepackage{xcolor}
\usepackage{pgfplots}
\pgfplotsset{compat=1.18}
\usepackage{caption}
\usepackage{subcaption}
\usepackage{balance}

\newcommand{\ie}{\textit{i.e.}}
\newcommand{\etc}{\textit{etc.}}

\renewcommand{\v}{\mathbf{v}}

\newcommand{\R}{\mathbb{R}}

\newcommand{\cossim}{\rm{cos}}

\newcommand{\per}{pEBR}

\title{pEBR: A Probabilistic Approach to Embedding Based Retrieval
}

\author{
  Han Zhang$^1$, \quad Yunjiang Jiang$^2$, \quad Mingming Li$^3$, \\
  \textbf{Haowei Yuan$^4$, \quad Yiming Qiu$^{1\dagger}$,  \quad Wen-Yun Yang}$^{1\dagger}$ \\
  $^1$JD.com, Beijing, China \qquad $^2$Meta, USA \qquad $^4$Shanghai Jiaotong University\\   \qquad $^3$Institute of Information Engineering, Chinese Academy of Sciences, Beijing, China \\
  \texttt{\{zhanghang33, qiuyiming3\}@jd.com} \qquad \texttt{limingming@iie.ac.cn} \\
  \texttt{\{yunjiangster, wenyun.yang\}@gmail.com}  \qquad \texttt{yhw@sjtu.edu.cn}
}
\begin{document}
\maketitle
\renewcommand{\thefootnote}{}
\footnotetext{$^\dagger$ Corresponding authors.}
\renewcommand{\thefootnote}{\arabic{footnote}}

\begin{abstract}
Embedding-based retrieval aims to learn a shared semantic representation space for both queries and items, enabling efficient and effective item retrieval through approximate nearest neighbor (ANN) algorithms. In current industrial practice, retrieval systems typically retrieve a fixed number of items for each query. However, this fixed-size retrieval often results in insufficient recall for head queries and low precision for tail queries. This limitation largely stems from the dominance of frequentist approaches in loss function design, which fail to address this challenge in industry. In this paper, we propose a novel \textbf{p}robabilistic \textbf{E}mbedding-\textbf{B}ased \textbf{R}etrieval (\textbf{pEBR}) framework. Our method models the item distribution conditioned on each query, enabling the use of a dynamic cosine similarity threshold derived from the cumulative distribution function (CDF) of the probabilistic model. Experimental results demonstrate that pEBR significantly improves both retrieval precision and recall. Furthermore, ablation studies reveal that the probabilistic formulation effectively captures the inherent differences between head-to-tail queries.


\end{abstract}

\section{Introduction}
Embedding retrieval is the core technology in search \citep{liu2021que2search, zhao2024dense,bao2024beyond} and recommendation\cite{youtube2016,li2020symmetric,xing2025esans}. 
The classic approaches are DSSM-like models~\citep{huang2013learning,zhang2020towards,qiu2022pre,li2021embedding,wang2023learning,bao2024beyond}
{\etc}, where the query and item are embedded into dense vectors in the same semantic space. The relevant items are then retrieved by calculating the cosine similarity between the query vectors and the item vectors.

However, we notice that existing works on embedding-based retrieval do not pay enough attention to the model training algorithm, especially the loss function, which in practice causes some tricky problems as follows: 
1) \textbf{Heuristic and suboptimal retrieval cutoffs.} Existing embedding-based retrieval systems rely on heuristic retrieval cutoffs, such as a fixed number of items or a fixed cosine similarity threshold. This leads to imbalanced retrieval performance: head queries suffer from low recall, while tail queries suffer from low precision. For instance, "iPhone 16" is a very specific query with few relevant items, while "gift" is broad, covering many products across categories. Thus, one can easily find that retrieving a fixed number of items would be suboptimal here, since if the number is small we will miss many relevant products for "gift" and if the number is large we will introduce many irrelevant items for "iPhone 16". On the other hand, it is also suboptimal to use a fixed cosine threshold, since all the existing model training algorithms have never taken that into account in their design. 
2) \textbf{Ignoring Query Popularity Bias.} Current two-tower training algorithms treat all queries equally and lack a probabilistic formulation to model item relevance distributions. As a result, they fail to account for query popularity differences (head vs. tail queries) and struggle to generalize, especially for long-tail queries with limited training data.

Before we introduce our new approach to address the above shortcomings, let us first review previous related works:
probabilistic models, or more precisely probabilistic graphical models, which have a long history of success in machine learning~\citep{murphy2012machine,koller2019book}. Several models have significantly changed the landscape of machine learning in recent decades, such as the Baum-Welch algorithm for training~\citep {mccallum2004hidden}, LDA~\citep{blei2003latent}, and probabilistic modeling with neural network \cite{nguyen2017mixture, malinin2019uncertainty,kohl2018probabilistic}.

Inspired by the previous successes of probabilistic spirits, here we propose a redesign of the embedding-based retrieval to address the above challenges by using probabilistic modeling.
Our contributions and the overview of the following sections can be summarized as follows.
\begin{itemize}
 \item We first formalize the conventional two-tower model with a frequentist perspective and systematically analyze its inherent limitations.
 \item We then propose two novel probabilistic methods:  one based on maximum likelihood estimation and another based on contrastive estimation; the latter includes two instances, ExpNCE and BetaNCE.
 \item We conduct both offline and online experiments to demonstrate the effectiveness of our model, and perform ablation studies to help understand how the model works.

\end{itemize}

\section{Problem and Limitations}
\label{sec:method}

In this section, we first formulate the embedding retrieval problem based on a two-tower architecture. Next, we review the existing frequentist approaches and present limitations. 

\subsection{Problem Definition and Notations}
\label{sec:definition}
The embedding retrieval model is typically composed of a query (or a user) tower and an item tower. For a given query $q$ and an item $d$, the scoring output of the model is: 
\begin{equation}
    f(q, d) = s(\v_q, \v_d),
\end{equation}
 where $\v_q \in \R^n$ and $\v_d \in \R^n$ 
denotes the  embedding of the query and the item, respectively. 
$s(.,.)$ computes the final score between the query and the item, such as the cosine similarity function, equivalently inner product between normalized vectors, \ie,  $s(\v_q, \v_d) = \v_q^\top \v_d$, where the superscript $^\top$ denotes matrix transpose.
The objective is to select the top $k$  items from a pool of candidate items for each given query $\v_q$.

\subsection{Existing Frequentist Approaches}
Most existing embedding retrieval adopt a frequentist approach to loss function design, which can be categorized into two types: point-wise loss and pair-wise loss. We discuss each category in detail.

\subsubsection{Point-Wise Loss}
\label{sec:pointwise}
The point-wise based method converts the original retrieval task into a binary classification, whose goal is to optimize the embedding space where the similarity between the query and its relevant item is maximized, while the similarity between the query and the irrelevant items is minimized. It usually adopts the sigmoid cross-entropy to train the model:
\resizebox{0.99\linewidth}{!}{
\begin{minipage}{\linewidth}
\begin{align*}
L_{\textrm{pointwise}}(\mathcal{D}) = - \sum_{i} \log \sigma(s(\v_{q_i}, \v_{d^+_i})) \\ + \sum_{ij} \log \sigma(s(\v_{q_i}, \v_{d_{ij}^{-}})),
\end{align*}
\end{minipage}
}
where $d^+_i$ denotes the relevant items for query  $q_i$ and $d_{ij}^{-}$ denotes the irrelevant ones, $\sigma(x) = 1 / (1 + \exp(-x))$ is the standard sigmoid function. 

\subsubsection{Pair-Wise Loss}
\label{sec:pairwise}
This kind of method aims to learn the partial order relationship between positive and negative items from the perspective of metric learning.
The classical work contains triple loss, margin loss, A-Softmax loss, and several variants
(margin angle cosine). 
Without loss of generality, we formulate the loss as softmax cross-entropy:
{\small
\begin{align*}
L_{\textrm{pairwise}}(\mathcal{D}) 
&= - \sum_{i} \log \left(\frac{\exp(s(\v_{q_i}, \v_{d^+_i})/\tau)}{\sum_j \exp(s(\v_{q_i}, \v_{d_{ij}})/\tau)} \right),
\end{align*}
}
where $\tau$ is the so-called temperature parameter: a lower temperature is less forgiving of mis-prediction of positive items by the model. In the same direction, researchers later proposed more advanced loss functions, by introducing max margin in angle space~\citep{liu2017cvpr}, in cosine space~\citep{cosface2018cvpr} and so on.

\subsubsection{Limitations}
Both point-wise and pair-wise loss functions have their advantages and limitations. Point-wise loss functions are relatively simpler to optimize, while they may not capture the partial order relationships effectively. In contrast, Pair-wise loss functions explicitly consider the relative ranking between items. As a result, pair-wise loss functions usually achieve better performance in embedding retrieval tasks.
While both of them are frequentist approaches, in the sense that no underlying probabilistic distribution are learned and consequently there is no cutoff threshold in principle when retrieving items. Thus, we propose the following probabilistic approach to a more theoretically well founded retrieval.

\section{Method}
 In this section, we explore the probabilistic approach, which contains the choice of the item distribution function and our carefully designed loss function based on the item distribution.
\subsection{Retrieval Embeddings as a Maximum Likelihood Estimator}
\label{sec:mle}
Given a query $q$, we propose to model the probability of retrieving item $d$: 
\begin{equation*}
        p(d|q) \propto p(r_{d,q}|q),
\end{equation*}
where $r_{d,q}$ represents the relevance between the query $q$ and document $d$. 
For all relevant items ${\textbf{d}^+}$, we assume $r_{d^+,q}$, where $d^+ \in \textbf{d}^+$, follows a distribution whose probability density function is $f_\theta$. For all irrelevant items ${\textbf{d}^-}$, we assume $r_{d^-,q}$, where $d^- \in \textbf{d}^-$, follows a distribution of whose probability density function is $h_\theta$. The likelihood function can be defined as
\begin{align*}
    L(\theta) & = \prod_{q}\left(\prod_{\textbf{d}^+} p(d^+|q)\prod_{\textbf{d}^-} p(d^-|q)\right) \\
              & \propto \prod_{q}\left(\prod_{\textbf{d}^+} p(r_{d^+, q}|q)\prod_{\textbf{d}^-} p(r_{d^-, q}|q)\right)
              \\
              & = \prod_{q}\left(\prod_{\textbf{d}^+} f_{\theta,q}(r_{d^+, q})\prod_{\textbf{d}^-} h_{\theta, q}(r_{d^-, q})\right).
\end{align*}

Importantly, the density for relevant items $p(r_{d^+, q}|q) = f_{\theta, q}(r_{d^+, q})$ and irrelevant items $p(d^-|q) = h_{\theta, q}(r_{d^-, q)}$  are both query dependent. This is a useful generalization from fixed density since different queries have different semantic scopes.
Finally the objective function can be defined as the log-likelihood:
{\small
\[
    l(\theta)
               = \sum_{q}\left(\sum_{\textbf{d}^+} \log f_{\theta, q}(r_{d^+, q}) + \sum_{\textbf{d}^-} \log h_{\theta, q}(r_{d^-, q})\right).
\]
}

When we choose different distributions for relevant and irrelevant items, the loss function can resemble a point-wise loss, which may lead to suboptimal performance compared to pair-wise loss functions. To address the limitation, we propose new probabilistic loss functions based on the principles of contrastive loss, which is a well-known pair-wise loss function.

\subsection{Retrieval Embeddings as a Noise Contrastive Estimator}
\label{sec:nce}
Apart from the above maximum likelihood estimator, the most widely used technique for retrieval model optimization is based on the InfoNCE loss~\citep{oord2018representation}, which is a form of noise contrastive estimator of the model parameter. 
In that setting, we need to choose two distributions: positive sample distribution $p(d^+ | q)$, and background (noise) proposal distribution $p(d)$, where they are related in theory, if we know the joint distribution of $d$ and $q$. But in practice, we can either treat them as separate or simply hypothesize their ratio as the scoring function $r(d, q):= p(d | q) / p(d)$, without knowing them individually. 
The loss we are minimizing is thus the following negative log-likelihood of correctly identifying the positive item within the noise pool
\begin{align}
L_r = - \sum_i \log \frac{r(d^+_i, q_i)}{\sum_j r(d_{ij}, q_i)}.
\label{eq:log_ll}
\end{align}

By minimizing the loss function, the model aims to maximize $p(d^+|q)$ for the query $q$ and one of its relevant items $d^+$, while pushing away $q$ and its irrelevant items $d^+_j$, thus assign higher similarities to relevant items compared to irrelevant items.
Based on the definition,  we propose two types of distributions as the basis of the estimator, the truncated exponential distribution ExpNCE and the Beta distribution BetaNCE.

\subsubsection{Parametric Exponential InfoNCE (ExpNCE)}
\label{sec:expnce}

Here we propose to use the following truncated exponential distribution density function as the scoring function $r(d,q) \propto \exp(\cossim(\v_d, \v_q) / \tau_q)$ where the function $\cossim$ stands for the cosine similarity between the two vectors and the temperature $\tau_q$ is query dependent. This offers an interesting alternative to the standard InfoNCE loss with log bi-linear distribution as
{\small
\[
    L_{\rm{ExpNCE}} = \sum_i \log \left(1 + \frac{\sum_{j} \exp(\cossim(\v_{q_i}, \v_{d_{ij}^-})/\tau_q)}{\exp(\cossim(\v_{q_i}, \v_{d_i^+}) / \tau_q)}\right).
\]
}

A nice property of the above probabilistic modeling is that we can use a simple cumulative density function (CDF) to decide the cutoff threshold. The CDF for the above truncated exponential distribution can be derived as follows.
\begin{align*}
\mathbb{P}_{\rm{ExpNCE}}(x < t) 
    & = \frac{\int_{-1}^t e^{x/\tau_q}  dx}{\int_{-1}^1 e^{x/\tau_q}  dx} \\
    & = \frac{e^{t/\tau_q} - e^{-1/\tau_q}}{e^{1/\tau_q} - e^{-1/\tau_q}},
\end{align*}
which can be easily computed.

\subsubsection{Parametric Beta InfoNCE (BetaNCE)}
\label{sec:betance}
We call a probability distribution compactly supported if its cumulative distribution function $F$ satisfies $F((-\infty, -x]) = 1 - F([x, \infty)) = 0$ for some $x > 0$. In other words, all the mass is contained in the finite interval $[-x, x]$. 
The best-known family of compact distributions in statistics is probably the Beta distributions, whose pdf are defined on $[0, 1]$. Since the cosine similarity used in the two-tower model has a value range of $[-1,1]$, we expand the definition range of Beta distributions to $[-1,1]$ and define its density as
\begin{align*}
f_{\alpha, \beta}(x) = \frac{\Gamma(\alpha + \beta)}{2\Gamma(\alpha) \Gamma(\beta)} \left(\frac{1+x}{2}\right)^{\alpha -1} \left(\frac{1 - x}{2}\right)^{\beta - 1}.
\end{align*}

An interesting special case is when $\alpha = \beta =1$, $F_{1, 1}(x) = \frac{x + 1}2
$, which gives the following uniform CDF on $[-1, 1]$.

One difficulty working with Beta distributions is that its CDF has no closed form, but rather is given by the incomplete Beta function\footnote{\url{https://www.tensorflow.org/api_docs/python/tf/math/betainc}}: $B(t; \alpha, \beta) = \int_0^t x^{\alpha - 1}(1-x)^{\beta - 1} dx$ is the incomplete Beta integral. The CDF for the above Beta distribution can be derived as follows.
{\small
\begin{align*}
\mathbb{P}_{\rm{BetaNCE}}(x < t) 
    &= \frac{\int_{-1}^t ((1 + x)/2)^{\alpha - 1}((1 - x)/2)^{\beta - 1} dx }{\int_{-1}^1 ((1 + x)/2)^{\alpha - 1}((1 - x)/2)^{\beta - 1} dx } \\
    &=\frac{\int_0^{\frac{t + 1}{2}} x^{\alpha - 1}(1 - x)^{\beta - 1} dx }{\int_0^1 x^{\alpha - 1}(1 - x)^{\beta - 1} dx } \\
    &= \frac{\Gamma(\alpha + \beta)}{\Gamma(\alpha) \Gamma(\beta)} B(\frac{t + 1}{2}; \alpha, \beta).
\end{align*}
}

Fortunately we only need the PDF during training; the CDF is needed only once, to determine the retrieval threshold after training (see Appendix~\ref{inference_truncation}). Then we assume $p(d|q)$ and $p(d)$ follow beta distributions like:
\begin{align*}
    p(d|q) &= g_q(s) \propto z^{\alpha_g(q)} (1 - z))^{\beta_g(q)},
\end{align*}
\begin{align*}
    p(d) &= k_q(s) \propto z^{\alpha_k(q)} (1 - z))^{\beta_k(q)},
\end{align*}
where $z=\frac{1 + s}{2}$, s is the cosine similarity between the query and item, $\alpha_g(q)$ and $\beta_g(q)$ are encoders based on the query representation to distinguish the learning of item distribution, and so do $\alpha_k(q)$ and $\beta_k(q)$.
According to the definition of $r(d, q)$ in Section~\ref{sec:nce}, we can get
\begin{align*}
    r(d, q) = \frac{g_q(s)}{k_q(s)} \propto z^{\alpha_g(q) - \alpha_k(q)} (1 - z)^{\beta_g(q) - \beta_k(q)},
\end{align*}
which is again a Beta density. Note that $\alpha, \beta$ correspond to the number of successes and failures that lead to the Beta distribution. We can view both $p(d|q)$ and $p(d)$ as having the same number of failures (being a negative document for $q$), but $p(d|q)$ has some additional successes. Thus we can hypothesize that $\beta_g = \beta_k$, and therefore get the final BetaNCE loss of log-likelihood in Equation (~\ref{eq:log_ll}) as
{\small
\begin{align*}
    L_{\rm{BetaNCE}} &= - \sum_i \log \left(\frac
        {z(\v_{q_i}, \v_{d_i^+})^{\alpha_g(q) - \alpha_k(q)}}
        {\sum_j z(\v_{q_i}, \v_{d_{ij}})^{\alpha_g(q) - \alpha_k(q)}}
    \right) \\
    &=- \sum_i \log \left(\frac
        {e^{\log z(\v_{q_i}, \v_{d_i^+}) / \tau_q}}
        {\sum_j e^{\log z(\v_{q_i}, \v_{d_{ij}}) / \tau_q}}
    \right),
\end{align*}
}
where $\tau_q = \left(\alpha_g(q) - \alpha_k(q)\right)^{-1}$. Thus compared with InfoNCE, the main difference is the logarithmic transformation applied to the cosine similarity. Since the cosine similarity $s$ is empirically always bounded away from $-1$, the logarithm presents no numerical difficulty.

After the training, we can get back the Beta distribution parameters from the learned $\tau_q$ by fixing the background distribution parameters $\alpha_k(q) \equiv 1$, $\beta_k(q) \equiv 1$, that is, the uniform distribution on the unit interval:
\begin{align}
\label{eq:betance_beta}
\alpha_g (q) = \tau_q^{-1} \text{ and } \beta_g(q) = 1.
\end{align}

\section{Experiments}

\begin{table*}[t]
\centering
\small
\caption{Comparison of retrieval quality between our proposed method and baseline methods.}
\label{tab:comparison}
\centering
\begin{tabular}{c|cc|cc|cc|cc}
\hline
           &  \multicolumn{2}{c|}{All Queries}       &  \multicolumn{2}{c|}{Head Queries}       & \multicolumn{2}{c|}{Torso Queries}     & \multicolumn{2}{c}{Tail Queries} \\ 
           &  P@1500 & R@1500 &  P@1500 & R@1500 & P@1500 & R@1500 & P@1500 & R@1500           \\ \hline\hline
DSSM-topk  &   0.327\%    &    93.29\%   &   0.782\%    &    90.38\%   &   0.211\%    &    94.36\%    &   0.126\%   & 94.36\% \\
DSSM-score &   0.435\%    &   93.64\%    &   0.841\%    &    90.77\%   &   0.339\%    &    94.64\%    &   0.275\%   & 94.57\% \\
\per &   \textbf{0.583}\%    &   \textbf{94.08}\%    &   \textbf{0.945}\%    &    \textbf{91.42}\%   &   \textbf{0.513}\%    &    \textbf{95.00}\%    &   \textbf{0.450}\%   & \textbf{94.73}\% \\ \hline
\end{tabular}
\end{table*}

\begin{table}[t]
  \caption{Online A/B testing.}
  \label{online}
  \centering
  \resizebox{0.85\columnwidth}{!}{
  \begin{tabular}{ccc}
    \cline{1-3}
     & UCVR & UCTR \\
     \hline
     \hline
    Relative improvement & \textbf{+0.190\%} & \textbf{+0.093\%} \\
    P-value & \textbf{0.0183} & \textbf{0.0213} \\
    \cline{1-3}
  \end{tabular}}
\end{table}

\label{sec:experiment}
In this section, we first explain the details of the dataset, experimental setup, baseline methods and evaluation metrics. Then we show the comparison of experimental results between the baseline methods and our proposed method {\per}. Finally we show the ablation study to intuitively illustrate how {\per} could perform better.

\subsection{Experimental Setup}
\subsubsection{Dataset and Setup} 
Our model was trained on a randomly sampled dataset consisting of 87 million user click logs collected over a period of 60 days. We trained the model on an Nvidia GPU A100 and employed the AdaGrad optimizer with a learning rate of 0.05,  batch size of 1024, and an embedding size of 128. The training process converged after approximately 200,000 steps, which took around 12 hours.

\subsubsection{Comparative Backbone}
Our method is compatible with most two-tower retrieval models since it primarily modifies the loss function rather than the model architecture. Thus, we choose a classic two-tower  DSSM~\citep{huang2013learning,qiu2022pre} as the backbone model. Then, we focus on cutting off items retrieved by the DSSM model with three methods: 
\begin{itemize}
    \item \textbf{DSSM-topk} refers to the method that cuts off items using a fixed threshold of number. 
    \item \textbf{DSSM-score} refers to the method that cuts off items with a fixed threshold of relevance score. Items with relevance scores below the threshold are discarded.
    \item \textbf{{\per}} refers to our proposed probabilistic method which cuts off items with a threshold derived by a probabilistic model, specifically the CDF value of the learned item distribution. We are using a default CDF value $0.5$ if not further specified. In this experiment, we focus on using BetaNCE with $\beta=1$ as shown in Equation (\ref{eq:betance_beta}) to demonstrate the effectiveness of the methods. But our method is general enough to accommodate other distributions, such as  ExpNCE.
\end{itemize}

As our method is model-agnostic and can be applied to any two-tower architecture, we also conduct experiments with a BGE~\cite{xiao2024c} (transformer-based) backbone to verify this compatibility. Details of these additional experiments are presented in Appendix~\ref{backbone_exp}.


\subsubsection{Experimental Metrics}
We use two widely reported metrics, Recall@$k$ and Precision@$k$, to evaluate the quality and effectiveness of retrieval models~\citep{zhang2021joint,zhu2018learning,lmmyuan2023multi}.  
There are some nuances that need a little more explanation. In DSSM-topk, a fixed number $k$ of items is retrieved for each query to calculate R@$k$ and P@$k$. While in DSSM-score and {\per}, a threshold of relevance score or CDF value is used to cut off retrieved items, which results in a variable number of items for each query. To ensure a fair comparison, we slightly adjust the threshold so that the average number of retrieved items across all queries equals $k$ when computing the evaluation metrics.

Moreover,  there is a tradeoff between precision and recall when varying the $k$ value. In the experiments, we choose $k=1500$ to optimize for recall, since it is the main goal of a retrieval system. Thus, the precision value is relatively low.


\begin{figure*}[t]
\centering
    \begin{subfigure}{0.25\textwidth}
        \centering
        \includegraphics[width=0.99\textwidth]{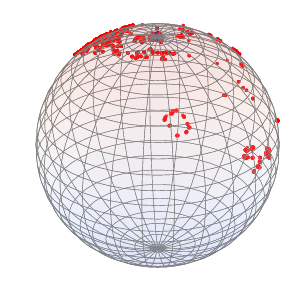}
        \caption{``phone''}
        \label{fig:head_query}
    \end{subfigure}
    \begin{subfigure}{0.25\textwidth}
        \centering
        \includegraphics[width=0.99\textwidth]{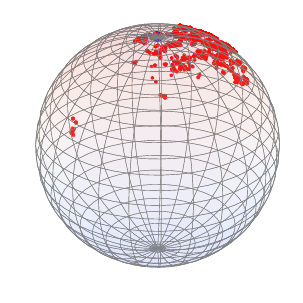}
        \caption{``Huawei phone 5g''}
        \label{fig:mid_query}
    \end{subfigure}
    \begin{subfigure}{0.25\textwidth}
        \centering
        \includegraphics[width=0.99\textwidth]{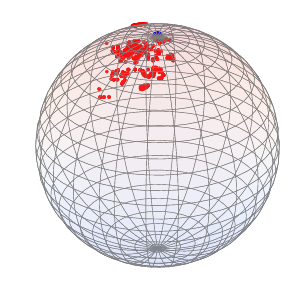}
        \caption{``Huawei mate50pro''}
        \label{fig:tail_query}
    \end{subfigure}
\caption{Relevant item distributions projected to a sphere for queries in head, torso, and tail categories, respectively.}

\label{fig:sphere}
\end{figure*}

\subsection{Experimental Results}
\label{sec:comparison}
Since retrieval performance varies considerably across head, torso, and tail queries, 
we adopt a tripartite dataset partitioning strategy for stratified evaluation.
As shown in Table~\ref{tab:comparison}, we can draw the following conclusions:
\begin{itemize}
    \item DSSM-score performs better than DSSM-topk on both R@1500 and P@1500, which is as expected since the retrieved items of DSSM-score have better relevance scores than DSSM-topk, because items with lower relevance scores are cut off.
    \item {\per} outperforms both DSSM-topk and DSSM-score, because
{\per} learns varying item distributions for different queries, allowing for the determination of dynamic and optimal relevance score thresholds, thus leading to enhanced overall retrieval performance. 
    \item {\per} achieves better performance on separated evaluation sets, {\ie} head queries, torso queries and tail queries. Moreover, we observe that the amount of improvement achieved by {\per} varies systematically with respect to query popularities. Specifically, compared to DSSM-topk, {\per} achieves improvements on head queries, torso queries and tail queries respectively. This is because head queries normally have many more relevant items than tail queries; thus, a dynamic cutoff threshold could benefit head queries to retrieve more items significantly.  What's more, tail queries normally do not have many relevant items, typically just tens of items. Thus, retrieving a fixed number $k$ of items hurts the precision significantly. {\per} appears to solve this problem nicely.
\end{itemize}

\subsection{Online A/B Testing}
To assess the effectiveness of the proposed method in a real-world scenario, we performed comprehensive online A/B testing within an e-commerce search engine with hundreds of millions of users and billions of products. 

Through a continuous week-long controlled experiment, table \ref{online} shows that our proposed method achieved consistant improvements over the previous baselines, yielding a  0.19\% increase in User Conversion Rate (UCVR) (p-value < 0.02), translating into substantial revenue gains for the company.

\begin{figure*}[!t]
\centering
\includegraphics[width=480pt]{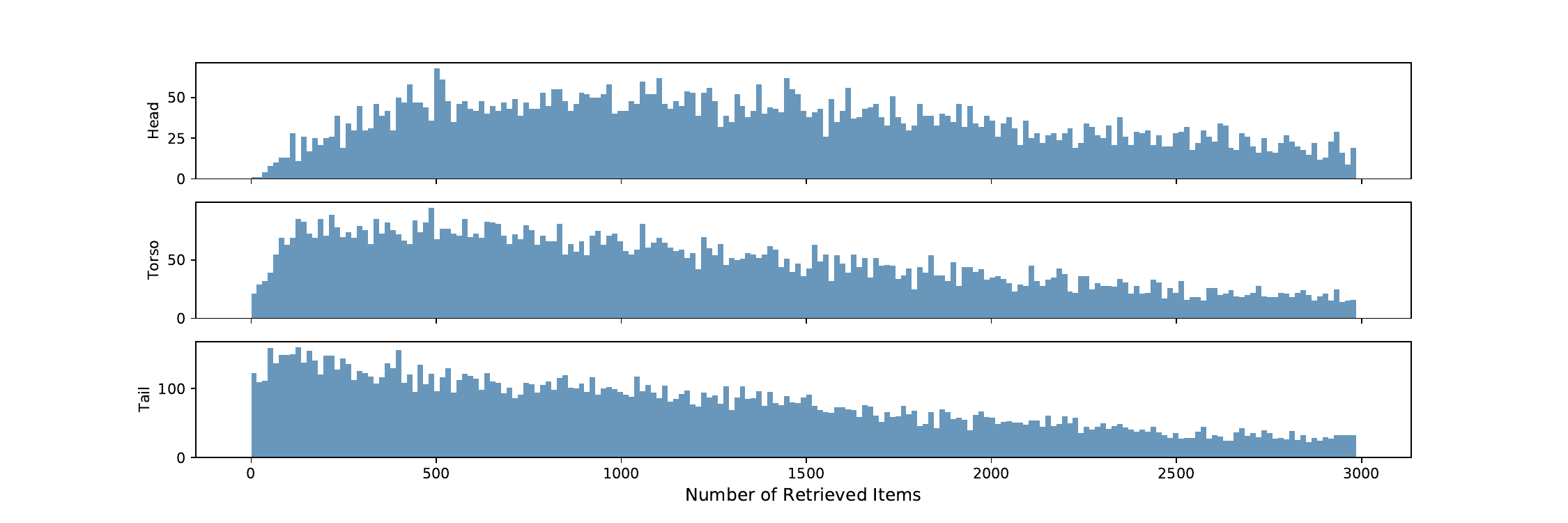}
\caption{Histogram of the retrieved item numbers for head, torso, and tail queries, with the CDF value set to 0.985.}
\label{fig:dynamic_cutoff}
\end{figure*}
\begin{table*}[!t]
    \centering
    \caption{Average numbers of retrieved items for different cutoff CDF values.}
    \label{tab:dynamic_cutoff}
    \begin{tabular}{c|rrrrrrrr}
    \hline
        Cutoff CDF Value & 0.99 & 0.95 & 0.9 & 0.8 & 0.7 & 0.6 & 0.5 & 0.4 \\ 
    \hline
    \hline
        Head Queries   & 2253.12 & 1443.51 & 1073.79 & 761.06 & 600.31 & 441.51 & 227.42 & 66.36 \\ 
        Torso Queries  & 1897.53 & 961.90 & 617.27 & 334.10 & 200.17 & 111.28 & 48.37 & 12.89 \\ 
        Tail Queries   & 1804.80 & 813.24 & 473.40 & 215.50 & 113.99 & 58.86 & 21.46 & 4.60 \\ 
    \hline
    \end{tabular}
\end{table*}
\subsection{Ablation Study}
\label{sec:ablation}
\subsubsection{Visual Illustration of Distributions}
As shown in Figure~\ref{fig:sphere}, we visualize the relevant item distribution for three queries with different frequencies.  We first need to apply a dimension reduction technique, specifically t-SNE~\citep{van2008visualizing} here to reduce the dimension to 3-D. Then we normalize the vectors as unit vectors and plot them on a sphere. 
The head query, ``phone'', is a quite general one that can retrieve phones of various brands and models. As a result, the retrieved items are widely distributed and dispersed across the surface of the sphere. The torso query ``Huawei phone 5g'' is a more specific one as it focuses on phones from the brand Huawei and with 5G capability. Consequently, the item distribution is more concentrated and has a narrower scope compared to the query ``phone''. The tail query, ``Huawei mate50pro'', is highly specific as it specifies the brand (Huawei) and model (mate50pro), thus the number of relevant items is very small and the distribution becomes even more concentrated. These differences in item distributions reaffirm the conclusion that cutting off retrieved items by a fixed threshold of item numbers or a fixed relevance score is suboptimal for embedding retrieval.

\subsubsection{Dynamic Retrieval Effect}
To investigate the effect of dynamic retrieval, we
analyze the retrieval results from two perspectives.
On the one hand, we examine the distribution of the
number of retrieved items under a fixed probabilistic CDF threshold. On the other hand, we evaluated
the number of items recovered under different probabilistic CDF thresholds. These analyses allow us to
gain insight into how dynamic retrieval influences
the retrieval results.

In Figure \ref{fig:dynamic_cutoff}, we show the histogram of retrieved
item numbers cut off by the CDF value 0.985 for
both head, torso, and tail queries. In detail, we
filter out the items with the cosine similarity x
bellowing a threshold t to ensure the equation
P(x >= t) = 0.985. In general, head queries
have a more uniform distribution, and the numbers
lie mostly in the range of [500, 1500], while the torso
and tail queries share a similar skewed distribution
and the numbers lie in the range of [0, 1000] and [0,
500], respectively. This indicates that head queries
retrieve more items than torso and tail queries,
which confirms the assumption that queries with higher popularity need more candidates and queries with lower popularity need fewer.

In Table \ref{tab:dynamic_cutoff}, we present the average number of
retrieved items under different CDF values with
the same cutting off strategy in Figure \ref{fig:dynamic_cutoff}. As the
probability decreases, queries claim a higher relevance level for items, and thus the average number of
retrieved items decreases accordingly. Comparing
different queries, head queries retrieve more items
than torso and tail queries under all CDF thresholds,
which again confirms the conclusion above.

\section{Conclusion}
In this paper, we have proposed a novel probabilistic embedding based retrieval model, namely {\per}, to address the challenges of insufficient retrieval for head queries and irrelevant retrieval for tail queries. 
Based on the noise constrastive estimator, we have proposed two instance models: ExpNCE and BetaNCE, with the assumption that relative items follow truncated exponential distribution or beta distribution, which allow us to easily compute a probabilistic CDF threshold instead of relying on fixed thresholds, such as a fixed number of items or a fixed relevance score. 
Comprehensive experiments and ablation studies show that the proposed method not only achieves better performance in terms of recall and precision metrics, but also present desirable item distribution and number of retrieved items for head and tail queries.

\section{Limitations}
While our probabilistic model demonstrates significant improvements in retrieval accuracy and system performance, it remains inherently constrained by the two-tower and late-interaction retrieval paradigms (e.g., DSSM architectures). Notably, the emerging paradigm of generative retrieval \cite{lmmkuai2024breaking,lmmli2024generative,deng2025onerecunifyingretrieverank} powered by large language models (LLMs) introduce new opportunities and challenges that our current framework does not address.
We identify two promising directions for future research: 1) developing hybrid architectures that integrate the robustness and reliability of our probabilistic modeling with the semantic flexibility of generative retrieval, particularly for dynamic cutoff optimization. 2) Exploring constrained generation techniques that retain the precision and advantages of our current approach while mitigating LLM hallucinations through adaptive distribution modeling, thereby enhancing both accuracy and reliability in retrieval tasks.

\section{Acknowledgments}
We would like to express our sincere gratitude to Lin Liu, Songlin Wang, and Sulong Xu for their valuable guidance and technical support throughout this research. Their insightful suggestions and continuous encouragement were instrumental in shaping the direction of our work.
\bibliographystyle{acl_natbib}
\bibliography{reference}

\clearpage
\appendix

\section{Inference Truncation} \label{inference_truncation}
The major goal of using query dependent item distribution in neural retrieval is to give statistical meaning to the retrieved candidate set. Previously a somewhat arbitrary combination of cosine similarity threshold and top-K threshold are used. 
\begin{align*}
S_{t, K}(q) = \rm{Top}_K^{\cossim(q, \cdot)}\left(\{d_i: \cossim(q, d_i) \ge t\}\right)
\end{align*}
The cosine threshold does not account for the variability of item distributions across different queries, while the top-K threshold is mainly to save inference cost. 

Now for a given $q$-isotropic spherical distribution $H$, whose marginal density with respect to $\cossim(q, \cdot)$ is given by $h: [-1, 1] \to \R_+$, we can compute its CDF as follows
\begin{align} \label{cumulative_probability}
\mathbb{P}(H < t) = \frac{\int_{-1}^t h(x) (1 - x^2)^{(n-3)/2} dx }{\int_{-1}^1 h(x) (1 - x^2)^{(n-3)/2} dx}.
\end{align}

For a given $h$ and $t$, \eqref{cumulative_probability} can be computed numerically. For the two special $h$ we are concerned here, we can give semi-closed forms:
\begin{itemize}
\item For the Beta density (BetaNCE) $h(x) \propto (1 + x)^{\alpha - 1} (1 - x)^{\beta -1}$, the normalization constant, after the rescaling $[-1,1] \to [0, 1]$, is a complete Beta integral

\resizebox{\linewidth}{!}{
\begin{minipage}{\linewidth}
\begin{align*}
\int_0^1 x^{\alpha + \frac{n-5}{2}} ( 1 -x)^{\beta + \frac{n-5}{2}} dx &= B(\alpha + \frac{n-3}{2}, \beta + \frac{n-3}{2})\\
&= \frac{\Gamma(\alpha + \frac{n-5}{2})\Gamma(\beta + \frac{n-5}{2})}{\Gamma(\alpha + \beta + n - 4)},
\end{align*}
\end{minipage}
}

while the numerator is proportional to an incomplete Beta integral
\resizebox{\linewidth}{!}{
\begin{minipage}{\linewidth}
\begin{align*}
\int_0^{\frac{1+t}{2}} x^{\alpha + \frac{n-5}{2}} ( 1 -x)^{\beta + \frac{n-5}{2}} dx =: B_{\frac{1+t}{2}}(\alpha + \frac{n-3}{2}, \beta + \frac{n-3}{2}).
\end{align*}
\end{minipage}
}

\item For the truncated exponential density (corresponding to InfoNCE), $h(x) \propto e^{x / \tau}$, the integral we need to compute is the following modified Bessel integral
\resizebox{\linewidth}{!}{
\begin{minipage}{\linewidth}
\begin{align*}
\int_{-1}^t e^{x/\tau} (1 -x^2)^{\frac{n-3}{2}} dx = \int_{-1}^t e^{x/\tau}(1 + x)^{\frac{n-1}{2} - 1} (1 - x)^{\frac{n-1}{2} - 1} dx.
\end{align*}
\end{minipage}
}
This admits a closed form solution for any $t$ provided $n > 2$ is odd, however the solution has alternating signs, which is numerically unstable especially for large $n$.
\end{itemize}
Due to the difficulty of exact solutions, we resort to numeric quadratures:

\begin{verbatim}

import scipy.integrate as integrate
import scipy.special as special
import math

def BetaInt(alpha, beta, t):
    return integrate.quad(
        lambda x: (1+x) ** (alpha-1) * \
            (1-x) ** (beta-1), -1, t)

def BetaExpInt(alpha, beta, tau, t):
    return integrate.quad(
        lambda x: (1+x) ** (alpha-1) * \
            (1-x) ** (beta-1) * \
            math.exp(x/tau), -1, t)

cache = {}

def CosineInvCDF(p, quad_fn, *args):
    values = cache.setdefault(
        quad_fn, {}).get(tuple(args))
    if not values:
        denom = quad_fn(*args, 1)[0]
        values = cache[quad_fn][tuple(args)] = \
            [quad_fn(*args, i/500-1)[0] / denom \
                for i in range(1001)]
    if p >= values[-1]:
        return 1
    if p <= values[0]:
        return -1
    right = min(
        i for i, v in enumerate(values) if v >= p)
    left = right - 1
    return (right * (p-values[left]) + left *
        (values[right]-p)) / \
        (values[right]-values[left])/500 - 1


def BetaInvCDF(p, alpha, beta):
    return CosineInvCDF(
        p, BetaInt, alpha, beta)

def BetaExpInvCDF(p, alpha, beta, tau):
    return CosineInvCDF(
        p, BetaExpInt, alpha, beta, tau)

def InfoNCEInvCDF(p, n, tau):
    return CosineInvCDF(
        p, BetaExpInt, (n-1)/2, (n-1)/2, tau)

\end{verbatim}

\section{Backbone Experiments}
\label{backbone_exp}
Our proposed pEBR method is orthogonal to the underlying two-tower architecture and can be seamlessly integrated into various backbone models. To further validate the generalizability of our method, we conducted additional experiments using BGE, a transformer-based embedding model, as the backbone for our two-tower setup. As shown in table~\ref{tab:backbone_comparison}, our method consistently outperforms both top-K and fixed-score truncation approaches in terms of Recall@1500 across different backbone models, demonstrating the generalizability and effectiveness of pEBR.

\begin{table}[htbp]
  \centering
  \caption{Recall@1500 on different methods.}
  \resizebox{\columnwidth}{!}{
      \begin{tabular}{l|c|ccc}
        \hline
         & \textbf{All Queries} & \textbf{Head} & \textbf{Torso} & \textbf{Tail} \\
        \hline
        BGE-topk & 0.9592 & 0.9556 & 0.9622 & 0.9585 \\
        BGE-score & 0.9352 & 0.9415 & 0.9320 & 0.9347 \\
        BGE-pEBR & 0.9691 & 0.9730 & 0.9731 & 0.9641 \\
        \hline
      \end{tabular}
    }
  \label{tab:backbone_comparison}
\end{table}

\section{Different Corpus Sizes}
We have conducted supplementary experiments with varying training data sizes. The tabel~\ref{tab:corpus_size_comparison} shows that while absolute recall@1500 decreases as training data is reduced, the relative improvement brought by our method remains stable. This suggests that the query distribution modeling enabled by pEBR converges efficiently, and our method is robust even with moderate-sized datasets.

\begin{table}[htbp]
  \centering
  \caption{Recall@1500 on different corpus sizes.}
  \resizebox{0.8\columnwidth}{!}{
      \begin{tabular}{l|ccc}
        \hline
         & \textbf{87M} & \textbf{40M} & \textbf{20M} \\
        \hline
        BGE-topk & 0.9592 & 0.9535 & 0.9482 \\
        BGE-score & 0.9352 & 0.9153 & 0.9102 \\
        BGE-pEBR & 0.9691 & 0.9672 & 0.9524 \\
        \hline
      \end{tabular}
    }
  \label{tab:corpus_size_comparison}
\end{table}

\section{Different Recall Cutoffs}
To provide a more comprehensive evaluation, we also experimented with smaller k values (e.g., k=500 and k=1000). Detailed results are presented in the Table~\ref{tab:recall_cutoff_comparison}. 

\begin{table}[htbp]
  \centering
  \caption{Performance at different recall cutoffs.}
  \resizebox{0.85\columnwidth}{!}{
      \begin{tabular}{l|ccc}
        \hline
         & \textbf{R@1500} & \textbf{R@1000} & \textbf{R@500} \\
        \hline
        BGE-topk & 0.9592 & 0.9350 & 0.8833 \\
        BGE-score & 0.9352 & 0.8598 & 0.7344 \\
        BGE-pEBR & 0.9691 & 0.9589 & 0.9226 \\
        \hline
      \end{tabular}
    }
  \label{tab:recall_cutoff_comparison}
\end{table}




\end{document}